# On Hypercomplex Extensions of Quantum Theory

Special Issue - *Quantum Field Theory*

Daniel Sepunaru

RCQCE - Research Center for Quantum Communication,

Holon Academic Institute of Technology,

52 Golomb St., Holon 58102, Israel

**Abstract**

This paper discusses quantum mechanical schemas for describing waves with non-abelian phases, Fock spaces of annihilation-creation operators for these structures, and the Feynman recipe for obtaining descriptions of particle interactions with external fields.

**Introduction.**

Standard Hilbert space formulation of quantum theory provide a simple and convenient schema for describing phenomena involving electromagnetic interactions. Experimental evidence proving the existence of non-abelian gauge fields implies that a proper extension of this theory must exist.

In this paper, we consider the construction of consistent quantum mechanical frameworks (quantum mechanical descriptions of waves with non-abelian phases as a first step and a second quantization procedure as the next step) for describing non-abelian gauge fields. We illustrate the emerging structures employing the properties of one- and two-body states. Generalizations of Lorentz force and the derivation of corresponding non-abelian gauge fields according to the Feynman-Dyson schema is treated in the one-, three-, and seven-dimensional spaces of the internal parameters with special emphasis being placed upon the role of normed division (composition) algebras.

## The Hilbert spaces genealogical tree

In order to generate the extensions of functional analytical structures we use a sequence of composition algebras as the mathematical foundation of the theory, thereby obtaining a hierarchy which seems rich enough to incorporate existing experimental information about the known fundamental interactions.

Before starting it should be noted that use of composition algebras leads to grave restrictions: standard vector product multiplication exists only in vector spaces of dimensions 0, 1, 3, and 7 according to the solutions of the following relation (equation)[1],[2]:

$$n(n-1)(n-3)(n-7) = 0 \tag{1}$$

Beginning with the construction of single particle states, the following hierarchy structures with real scalar products exist:

a) Real valued state functions with a real scalar product – trivial:

$$(f,g)_R \equiv Tr(f,g) = f_0 g_0 = real \tag{2}$$

b) Complex valued state functions with a real scalar product:

$$f = f_0 e_0 + f_1 e_1 = complex$$

$$(f,g)_R \equiv Tr(f,g) = f_0 g_0 + f_1 g_1 = real \tag{3}$$

c) Quaternion valued state functions with a real scalar product:

$$f = f_0 e_0 + f_1 e_1 + f_2 e_2 + f_3 e_3 = \text{quaternion}$$

$$(f,g)_R \equiv Tr(f,g) = f_0 g_0 + f_1 g_1 + f_2 g_2 + f_3 g_3 = real, \tag{4}$$

where $f_0, f_i$ are functions over the reals with all the necessary properties required from the functional analysis; $e_i^2 = -e_0$, $\{e_i, e_j\} = 0$; $i \neq j$, $e_i e_0 = e_0 e_i = e_i$, $i,j = 1,2,3$; $e_0^2 = e_0$, $e_3 = e_1 e_2$.

That structure may also be generated by four dimension vectors:

$$(f,g) = Tr(f,g) - e_i Tr\{(f,g)e_i\}; \; i = 1,2,3$$

$$-e_1(f,g)e_1 = Tr(f,g) - e_1Tr\{(f,g)e_1\} + e_2Tr\{(f,g)e_2\} + e_3Tr\{(f,g)e_3\} \qquad (5)$$

$$-e_2(f,g)e_2 = Tr(f,g) + e_1Tr\{(f,g)e_1\} - e_2Tr\{(f,g)e_2\} + e_3Tr\{(f,g)e_3\}$$

$$-e_3(f,g)e_3 = Tr(f,g) + e_1Tr\{(f,g)e_1\} + e_2Tr\{(f,g)e_2\} - e_3Tr\{(f,g)e_3\}$$

The sum of Eqs.(5) gives us

$$(f,g)_R \equiv Tr(f,g) = \tfrac{1}{4}[(f,g) - e_i(f,g)e_i] = \tfrac{1}{4}[\bar{f}, -e_1\bar{f}, -e_2\bar{f}, -e_3\bar{f}] \cdot \begin{bmatrix} g \\ ge_1 \\ ge_2 \\ ge_3 \end{bmatrix} \qquad (6)$$

d) Octonion valued state functions over the reals with real scalar product:

$$f = f_i e_i = \text{octonion}; \quad i = 0,1,...,7$$

$$(f,g)_R \equiv Tr(f,g) = f_i g_i = real \qquad (7)$$

$$e_i^2 = -e_0, \quad \{e_i, e_j\} = 0; \, i \neq j, \, e_i e_0 = e_0 e_i = e_i, i, j = 1,2,...,7; \, e_0^2 = e_0,$$

$$e_i e_j = f_{ijk} e_k; \, i, j, k = 1,2,...7$$

$f_{ijk}$ is a completely antisymmetric seven-dimensional analog of the Levi-Civita symbol with the following multiplication table:

$$f_{ijk} = e_0; (i,j,k) = (123),(471),(257),(165),(624),(543),(736) \text{ for example.} \qquad (8)$$

Then

$$(f,g) = Tr(f,g) - e_i Tr\{(f,g)e_i\}; \, i = 1,...,7$$

$$-e_1(f,g)e_1 = Tr(f,g) + e_i Tr\{(f,g)e_i\} - 2e_1 Tr\{(f,g)e_1\}$$

$$-e_2(f,g)e_2 = Tr(f,g) + e_i Tr\{(f,g)e_i\} - 2e_2 Tr\{(f,g)e_2\}$$

$$-e_3(f,g)e_3 = Tr(f,g) + e_i Tr\{(f,g)e_i\} - 2e_3 Tr\{(f,g)e_3\}$$

$$-e_4(f,g)e_4 = Tr(f,g) + e_i Tr\{(f,g)e_i\} - 2e_4 Tr\{(f,g)e_4\}$$

$$-e_5(f,g)e_5 = Tr(f,g) + e_i Tr\{(f,g)e_i\} - 2e_5 Tr\{(f,g)e_5\}$$

$$-e_6(f,g)e_6 = Tr(f,g) + e_i Tr\{(f,g)e_i\} - 2e_6 Tr\{(f,g)e_6\}$$

$$-e_7(f,g)e_7 = Tr(f,g) + e_i Tr\{(f,g)e_i\} - 2e_7 Tr\{(f,g)e_7\}$$

so we obtain

$$(f,g)_R \equiv Tr(f,g) = \tfrac{1}{3}\bar{f}\cdot g + \tfrac{1}{12}[\bar{f},-e_1\bar{f},\ldots,-e_7\bar{f}]\cdot\begin{bmatrix} g \\ ge_1 \\ \ldots \\ \ldots \\ \ldots \\ ge_7 \end{bmatrix} \quad (9)$$

Matrix multiplication is performed here as with the usual associative algebras due to the validity of the Moufang identity [3]

$$(ax)(ya) = a(xy)a. \quad (10)$$

Now consider the sequence of structures generated by the complex scalar products:

e) Complex valued state functions with a complex scalar product – the standard mathematical formalism of non-relativistic quantum mechanics:

$$f = f_0 e_0 + f_1 e_1 = complex$$

$$(f,g)_C = \bar{f}\cdot g \quad (11)$$

The structure may also be generated by two dimensional vectors:

$$(f,g)_C \equiv Tr(f,g) - e_1 Tr\{(f,g)e_1\} = \tfrac{1}{2}[(f,g) - e_1(f,g)e_1] \quad (12)$$

or in the matrix notations

$$(f,g)_C = \tfrac{1}{2}[\bar{f},-e_1\bar{f}]\cdot\begin{bmatrix} g \\ ge_1 \end{bmatrix} \quad (13)$$

The group of transformations which leaves that complex scalar product invariant is $U(1)$.

f) Quaternion valued state functions with a complex scalar product

$$f = f_0 e_0 + f_1 e_1 + f_2 e_2 + f_3 e_3 = \text{quaternion}$$

Similarly to the previous procedure, the required complex scalar product may be generated by four dimension vectors:

$$(f,g) = Tr(f,g) - e_i Tr\{(f,g)e_i\}; \quad i = 1,2,3$$

$$-e_1(f,g)e_1 = Tr(f,g) + e_i Tr\{(f,g)e_i\} - 2e_1 Tr\{(f,g)e_1\}$$

The summation then give us:

$$(f,g)_C \equiv Tr(f,g) - e_1 Tr\{(f,g)e_1\} = \tfrac{1}{2}[(f,g) - e_1(f,g)e_1] \tag{14}$$

In matrix notations

$$(f,g)_C = \tfrac{1}{2}[\bar{f}, -e_1\bar{f}] \cdot \begin{bmatrix} g \\ ge_1 \end{bmatrix} \tag{15}$$

The group of transformations which leaves that complex scalar product invariant is $U(2)$:

$$f' \Rightarrow qfz \; ; \; |q| = 1; |z| = 1$$

$$(f',g')_C = (qfz, qgz)_C = \tfrac{1}{2}[\bar{z}(f,g)z - e_1\bar{z}(f,g)ze_1] = \bar{z}(f,g)_C z = (f,g)_C$$

g) Octonion valued state functions over the reals with complex scalar product:

$$f = f_i e_i = \text{octonion}; \; i = 0,1,\ldots,7$$

$$(f,g) = Tr(f,g) - e_i Tr\{(f,g)e_i\}; \; i = 1,2,\ldots,7$$

$$-e_1(f,g)e_1 = Tr(f,g) + e_i Tr\{(f,g)e_i\} - 2e_1 Tr\{(f,g)e_1\}$$

again we have

$$(f,g)_C \equiv Tr(f,g) - e_1 Tr\{(f,g)e_1\} = \tfrac{1}{2}[(f,g) - e_1(f,g)e_1] \tag{16}$$

and in matrix notations

$$(f,g)_C = \tfrac{1}{2}[\bar{f}, -e_1\bar{f}] \cdot \begin{bmatrix} g \\ ge_1 \end{bmatrix} \tag{17}$$

The group of transformations that leaves that complex scalar product invariant is $U(4)$.

h) Quaternion valued state functions with quaternion scalar product

$$f = f_0 e_0 + f_1 e_1 + f_2 e_2 + f_3 e_3 = \text{quaternion}$$

$$(f,g)_Q = \bar{f} \cdot g \tag{18}$$

i)  Octonion valued state functions with quaternion scalar product:

$$f = f_i e_i = \text{octonion}; \quad i = 0,1,...,7$$

$$f = \psi_1 + \psi_2 e_7; \quad g = \psi_3 + \psi_4 e_7;$$

$$\psi_1 = \varphi_0 e_0 + \varphi_1 e_1 + \varphi_2 e_2 + \varphi_3 e_3 = \text{quaternion}$$

$$\psi_2 = \varphi_7 e_0 + \varphi_4 e_1 + \varphi_5 e_2 + \varphi_6 e_3 = \text{quaternion}$$

$$\psi_3 = \chi_0 e_0 + \chi_1 e_1 + \chi_2 e_2 + \chi_3 e_3 = \text{quaternion}$$

$$\psi_4 = \chi_7 e_0 + \chi_4 e_1 + \chi_5 e_2 + \chi_6 e_3 = \text{quaternion}$$

where $\varphi_0, \varphi_i, \chi_0, \chi_i, i = 1,2,...,7$ are functions over the reals.

$$(f,g)_Q = [\overline{\psi}_1, -e_7\overline{\psi}_2] \cdot \begin{bmatrix} \psi_3 \\ \psi_4 e_7 \end{bmatrix}$$

$$= \overline{\psi}_1 \cdot \psi_3 - (e_7\overline{\psi}_2) \cdot (\psi_4 e_7) = \overline{\psi}_1 \cdot \psi_3 - e_7(\overline{\psi}_2 \cdot \psi_4) e_7 = \overline{\psi}_1 \cdot \psi_3 + \overline{\psi}_4 \cdot \psi_2 \qquad (19)$$

j)  Octonion valued state functions with octonion scalar product:

$$f = f_i e_i = \text{octonion}; \quad i = 0,1,...,7$$

$$(f,g)_O = \bar{f} \cdot g \qquad (20)$$

**Fock space in Hypercomplex Quantum Mechanics**

The next step is developing a second quantization procedure for our schema, for which an ideal gas consisting of identical particles is considered. Restricting the discussion to structures with a complex scalar product and giving the general procedure for the reduction of tensor product algebras, a suitable redefinition of the scalar products is obtained which allows the proper extension of the function analysis.

Let us consider the tensor product of N Hilbert spaces. The state is defined by

$$\Psi(f_1, f_2,..., f_N) \equiv f_1(x_1) \otimes f_2(x_2) \otimes ... \otimes f_N(x_N) \qquad (21)$$

In general Kronecker multiplication, an algebraic operation different from inner multiplication and which cannot be reduced to it, is used. It is distributive with the following properties:

$$(f_1 \otimes g_1) \cdot (f_2 \otimes g_2) = (f_1 \cdot f_2) \otimes (g_1 \cdot g_2) \tag{22}$$

$$Tr(f \otimes g) = Tr(f) \cdot Tr(g) \tag{23}$$

$$N(f \otimes g) = N(f) \cdot N(g) \tag{24}$$

$$\dim(f \otimes g) = \dim(f) \cdot \dim(g). \tag{25}$$

Therefore, the product of N Hilbert spaces has the dimension $2^N$. In the case when quaternions are used to describe a single particle state we obtain $4^N$ for the dimension of both the system states and the scalar products (and $8^N$ for the octonions correspondingly). System states in quantum theory are not observable quantities, therefore, we need not reduce their dimension. However, the scalar products are observable quantities and should be numbers belonging to the one of the composition algebras.

Tensor products in standard quantum mechanical theory should satisfy the following general requirements of a quantum mechanical system without interaction:

1) Each component of a tensor product is completely independent of the others.

2) Construction of tensor product spaces do not spoil the validity of the superposition principle in each space.

In order to satisfy the above conditions one needs two different units $e_2$ (for example, for the quaternion states) in the algebraic basis of the theory: one which does not commute with some other unit $e_1$ (these units are used for the description of the quantum mechanical state in the same space) and another which does commute with that same unit $e_1$ (this $e_2$ belongs to the second space and the quantum mechanical state in that space should be completely independent of the quantum mechanical state which belongs to the first space).

The use of the Kronecker multiplication leads to the validity of the superposition principle on the level of many-body states. These states again appear to be quantum mechanical states satisfying the basic principles of quantum mechanical theory.

Let us now consider an obvious example of waves with non-abelian phases: quaternion quantum mechanics with a complex scalar product. On the level of one-body theory, the quantum mechanical state is described by the following matrix representation :

$$\Psi_C(f) = \tfrac{1}{\sqrt{2}} \begin{pmatrix} f \\ fe_1 \end{pmatrix} \qquad (26)$$

$$f = f_0 e_0 + f_1 e_1 + f_2 e_2 + f_3 e_3$$

Superpositions are defined by

$$\Psi_C(fq_1 + gq_2) = \tfrac{1}{\sqrt{2}} \begin{pmatrix} fq_1 + gq_2 \\ (fq_1 + gq_2)e_1 \end{pmatrix} = \tfrac{1}{\sqrt{2}} \begin{pmatrix} fq_1 \\ fq_1 e_1 \end{pmatrix} + \tfrac{1}{\sqrt{2}} \begin{pmatrix} gq_2 \\ gq_2 e_1 \end{pmatrix} = \Psi_C(fq_1) + \Psi_C(gq_2) \qquad (27)$$

Superpositions are linear only with respect to complex numbers. This form generates a complex scalar product defined by (15).

The following form represents two-body state:

$$\Psi_C(f_1, f_2) = \tfrac{1}{2} \begin{pmatrix} f_1 \otimes f_2 \\ f_1 e_1 \otimes f_2 \\ f_1 \otimes f_2 e_1 \\ f_1 e_1 \otimes f_2 e_1 \end{pmatrix} \qquad (28)$$

The form of the three-body states and so on is obvious.

The quaternionic units are non-commuting and it is clear that only use of Kronecker products (direct product algebras) allows us to satisfy the conditions for construction of the many-body states.

Using (28) we have:

$$(\Psi_C(f_1, f_2), G_C(g_1, g_2)) = \tfrac{1}{4} \{(f_1, g_1)_C \otimes (f_2, g_2)_C\} \qquad (29)$$

That reduces the scalar product algebra to its subalgebra with the basis

$$1 \otimes 1, \quad e_1 \otimes e_1, \quad e_1 \otimes 1, \quad 1 \otimes e_1$$

Further reduction is achieved through introduction of the projection operators

$$Z_0^{(2)} = \tfrac{1}{2}(1 \otimes 1 - e_1 \otimes e_1) \qquad (30)$$

$$Z_1^{(2)} = \tfrac{1}{2}(e_1 \otimes 1 + 1 \otimes e_1)$$

It is easy to verify that

$$Z_0^2 = Z_0, \qquad Z_1^2 = -Z_0, \qquad Z_0 \cdot Z_1 = Z_1 \cdot Z_0 = Z_1.$$

Finally, the required redefinition is obtained

$$(\Psi_C(f_1, f_2), G_C(g_1, g_2))_C = e_0 Tr\{(f_1, g_1)_C \otimes (f_2, g_2)_C \otimes Z_0^{(2)}\} - e_1 Tr\{(f_1, g_1)_C \otimes (f_2, g_2)_C \otimes Z_1^{(2)}\} \quad (31)$$

Using (23) we have

$$(\Psi_C(f_1, f_2), G_C(g_1, g_2))_C = \tfrac{1}{4}\{(f_1, g_1)_C \cdot (f_2, g_2)_C\} \qquad (32)$$

and because of the factorized form of the scalar product, realization of the second quantization procedure may be carried out analogously to the standard rules. There is no a priori connection between $e_0$ and $e_1$ which appear in front of the traces in the definition of the scalar product and the operators $Z_0$, $Z_1$ that are inside the scalar product.

In general, we introduce the following algebraic generalization of the complex scalar product:

$$(\Psi_C(f_1, f_2, ..., f_N), G_C(g_1, g_2, ..., g_N))_C \equiv e_0 Tr\{(\Psi_C, G_C) \otimes Z_0^{(N)}\} - e_1 Tr\{(\Psi_C, G_C) \otimes Z_1^{(N)}\} \qquad (33)$$

where $e_0^2 = e_0; e_0 e_1 = e_1 e_0 = e_1; e_1^2 = -e_0$ and $Z_0^{(N)}$, $Z_1^{(N)}$ form a complex subalgebra of the direct product algebras of the obtained construct.

In three-body case they are:

$$Z_0^{(3)} = \tfrac{1}{4}(1 \otimes 1 \otimes 1 - e_1 \otimes e_1 \otimes 1 - e_1 \otimes 1 \otimes e_1 - 1 \otimes e_1 \otimes e_1)$$

$$Z_1^{(3)} = \tfrac{1}{4}(e_1 \otimes 1 \otimes 1 + 1 \otimes e_1 \otimes 1 + 1 \otimes 1 \otimes e_1 - e_1 \otimes e_1 \otimes e_1)$$

and so on. $\{1, e_1\}$ is the label for the generators of the complex field in each space.

The complex linear operators have the following form:

$$A_z = \begin{pmatrix} a_{11} & a_{12} \\ -a_{12} & a_{11} \end{pmatrix}$$

where matrix elements $a_{ij}$ are c-number operators over quaternions and in turn are assumed to be at least z-linear operators. The quaternion linear operators have the form

$$A_q = \begin{pmatrix} a_{11} & 0 \\ 0 & a_{11} \end{pmatrix}$$

where $a_{11}$ is q-linear operator over quaternions. It has the following structure:

$$a_{11} = a_{11}^0 e_0 + a_{11}^1 e_1 + a_{11}^2 e_2 + a_{11}^3 e_3$$

where $a_{11}^i$ are real operators.

In occupation number representation the states of a system of fermions (in space 1, two-body case) are given by:

$$|0> = \begin{pmatrix} Z_0^{(2)} \\ 0 \\ Z_1^{(2)} \\ 0 \end{pmatrix} \qquad |1>^i = \begin{pmatrix} 0 \\ (e_i \otimes 1)Z_0^{(2)} \\ 0 \\ (e_i \otimes 1)Z_1^{(2)} \end{pmatrix} ; \qquad i = 1,2,3 \qquad (34)$$

Then annihilation-creation operators have following form:

$$a_i = \tfrac{1}{2} \begin{pmatrix} 0 & -Z_0^{(2)}(e_i \otimes 1) & 0 & Z_1^{(2)}(e_i \otimes 1) \\ 0 & 0 & 0 & 0 \\ 0 & -Z_1^{(2)}(e_i \otimes 1) & 0 & -Z_0^{(2)}(e_i \otimes 1) \\ 0 & 0 & 0 & 0 \end{pmatrix} ; \qquad i = 1,2,3 \qquad (35)$$

$$a_i^+ = \tfrac{1}{2} \begin{pmatrix} 0 & 0 & 0 & 0 \\ (e_i \otimes 1)Z_0^{(2)} & 0 & -(e_i \otimes 1)Z_1^{(2)} & 0 \\ 0 & 0 & 0 & 0 \\ (e_i \otimes 1)Z_1^{(2)} & 0 & (e_i \otimes 1)Z_0^{(2)} & 0 \end{pmatrix} ;$$

$$a_i |0> = 0 ; \quad a_i^+ |0> = |1>^i ; \quad i, j = 1,2,3$$

$$a_i^+ |1>^j = 0; \qquad a_i |1>^i = |0> \qquad \text{(no summation)}; \qquad (36)$$

$$a_i |1>^j = 0; \qquad i \neq j$$

and thus we have almost canonical fermion commutation relations for the annihilation-creation operators:

$$\{a_i, a_i^+\} = 1 \quad \text{(no summation)} \quad i = 1,2,3 \tag{37}$$

The last example with a similar structure is octonionic quantum mechanics with a complex scalar product. Realization of this case occurs through matrix representation of the one body state (Eq.(26)):

$$\Psi_C(f) = \tfrac{1}{\sqrt{2}} \begin{pmatrix} f \\ fe_1 \end{pmatrix}$$

where $f = f_i e_i =$ octonion; $i = 0,1,...,7$.

The remaining construction is identical to the previous cases due to the Moufang identity (Eq.(10))

$$(ax)(ya) = a(xy)a \quad \forall a, x, y \text{ octonions}$$

Our case corresponding to the choice $a = e_1$; $e_1$ is a label for one of the octonionic units.

An annihilation-creation operators for a system of fermions have the unusual properties because of the following multiplication rule for octonions:

$$e_{i+3} e_7 = e_i \quad i = 1,2,3$$

Let us consider $|0>_1$, $|1>_1^i$ defined as in (34), where $i = 1,...7$ and $a_i, a_i^+$ defined as in (35), $i = 1,...7$.

Then

$$a_i |0>_1 = 0 \quad a_i^+ |0>_1 = |1>_1^i$$

$$a_i^+ |1>_1^j = 0 \quad a_i |1>_1^i = |0>_1 \quad i, j = 1,...7 \tag{38}$$

$$a_i |1>_1^j = 0 \quad \begin{array}{l} i \neq j; j \neq i+3, j > i \\ j \neq i-3, j < i \end{array}$$

But

$$a_{i+3} |1>_1^i = \begin{pmatrix} Z_1 \\ 0 \\ Z_0 \\ 0 \end{pmatrix} \equiv |0>_2 \tag{39}$$

$$a_i |1>_1^{i+3} = -|0>_2 \quad i = 1,2,3$$

and

$$a_i^+ |0>_2 = \begin{pmatrix} 0 \\ (e_i \otimes 1)Z_1 \\ 0 \\ -(e_i \otimes 1)Z_0 \end{pmatrix} \equiv |1>_2^i \qquad i = 1,\ldots,7 \qquad (40)$$

Then

$$a_i |0>_2 = 0 \qquad a_i^+ |1>_2^j = 0 \qquad i, j = 1,\ldots,7$$

$$a_i |1>_2^i = |0>_2 \qquad a_i |1>_2^j = 0 \qquad \begin{array}{l} j \neq i; j \neq i+3; j > i \\ j \neq i-3; j < i \end{array} \qquad (41)$$

and

$$a_i(a_i^+ |0>_{1,2}) + a_i^+(a_i |0>_{1,2}) = (a_i a_i^+)|0>_{1,2} + (a_i^+ a_i)|0>_{1,2} = |0>_{1,2}$$

$$\qquad\qquad i = 1,\ldots,7$$

$$a_i(a_i^+ |1>_{1,2}^i) + a_i^+(a_i |1>_{1,2}^i) = (a_i a_i^+)|1>_{1,2}^i + (a_i^+ a_i)|1>_{1,2}^i = |1>_{1,2}^i$$

$$a_i |1>_2^{i+3} = |0>_1 \ ; \qquad a_{i+3} |1>_2^i = -|0>_1 \ ; \qquad i = 1,2,3$$

$$a_i^+ (a_{i+3} |1>_1^i) = (a_i^+ a_{i+3})|1>_1^i = |1>_2^i \qquad (42)$$

$$a_i^+ (a_{i+3} |1>_2^i) = (a_i^+ a_{i+3})|1>_2^i = -|1>_1^i$$

$$\qquad\qquad i = 1,2,3$$

$$a_{i+3}^+ (a_i |1>_1^{i+3}) = (a_{i+3}^+ a_i)|1>_1^{i+3} = -|1>_2^{i+3}$$

$$a_{i+3}^+ (a_i |1>_2^{i+3}) = (a_{i+3}^+ a_i)|1>_2^{i+3} = |1>_1^{i+3}$$

All these quantum mechanical schemas shares common features: states that satisfy the z-linear superposition principle, scalar products are z-linear and the following theorem is valid (here only the two-body case is considered as generalization to other cases is obvious):

Beckett Theorem

$$(\Psi_C(f_1, f_2), G_C(g_1 z, g_2))_C = (\Psi_C(f_1, f_2), G_C(g_1, g_2 z))_C \qquad (43)$$

Proof:

$$(\Psi_C(f_1,f_2), G_C(g_1z, g_2))_C = e_0 Tr\{(f_1,g_1z)_C \otimes (f_2,g_2)_C \otimes Z_0^{(2)}\} - e_1 Tr\{(f_1,g_1z)_C \otimes (f_2,g_2)_C \otimes Z_1^{(2)}\}$$

$$= e_0 Tr\{\tfrac{1}{2}(f_1,g_1)_C \otimes (f_2,g_2)_C \otimes [a(1\otimes 1) + b(e_1 \otimes 1]\cdot[1\otimes 1 - e_1 \otimes e_1]\}$$
$$- e_1 Tr\{\tfrac{1}{2}(f_1,g_1)_C \otimes (f_2,g_2)_C \otimes [a(1\otimes 1) + b(e_1 \otimes 1]\cdot[e_1 \otimes 1 + 1\otimes e_1]\}$$

$$= e_0 Tr\{\tfrac{1}{2}(f_1,g_1)_C \otimes (f_2,g_2)_C \otimes [a(1\otimes 1) + b(1\otimes e_1]\cdot[1\otimes 1 - e_1 \otimes e_1]\}$$
$$- e_1 Tr\{\tfrac{1}{2}(f_1,g_1)_C \otimes (f_2,g_2)_C \otimes [a(1\otimes 1) + b(1\otimes e_1]\cdot[e_1 \otimes 1 + 1\otimes e_1]\} = (\Psi_C(f_1,f_2), G_C(g_1, g_2 z))_C$$

In general

$$(\Psi_C(f_1,f_2,...,f_N), G_C(g_1,g_2,...,g_i z,..., g_j,..., g_N))_C = (\Psi_C(f_1,f_2,...,f_N), G_C(g_1,g_2,...,g_i,..., g_j z,... g_N))_C$$

expresses the statement that the observable quantities are given only in terms of their relative phases.

**Interactions**

Now we are able to study the particle interactions. Choosing the Feynman route to investigate the available options, we begin [4] with classical Newtonian equations of motion for single, isolated particles

$$m\ddot{x}_j = F_j(x,\dot{x},t); \quad j=1,2,3 \tag{44}$$

supplemented by Heisenberg (quantum) commutation relations

$$[x_j, x_k] = 0 \tag{45}$$

$$m[x_j, \dot{x}_k] = i\hbar \delta_{jk} \tag{46}$$

Then the charge moving in the given electromagnetic field exerts the Lorentz force

$$F_j(x,\dot{x},t) = eE_j(x,t) + e\varepsilon_{jkl}\dot{x}_k H_l(x,t), \quad (c=1) \tag{47}$$

where $E(x,t)$ and $H(x,t)$ are defined by the Maxwell equations

$$div H = 0 \tag{48}$$

$$\frac{\partial H}{\partial t} + curl E = 0. \tag{49}$$

The particle <u>internal</u> parameters are described by the charge and current densities $\rho$, $j$, which are responsible for coupling with the external electromagnetic field. They are defined by the additional pair of the Maxwell equations:

$$div E = 4\pi\rho \tag{50}$$

$$\frac{\partial E}{\partial t} - curl H = -4\pi j \tag{51}$$

and

$$\frac{\partial \rho}{\partial t} + div j = 0 \tag{52}$$

The presence of that conservation law tell us that we have deal with additional internal symmetry of the system.

Now let us return to Equation (1). If we use conventional vector product multiplication, then the space dimensions are fixed by the roots of that equation

$$n = 1, \ n = 3 \text{ or } n = 7.$$

There are always only three dimensions in the real (outer) world with no experimental evidence whatsoever contradicting this premise. Inner space, however, has a dimension of $n = 1$. Vector products in the inner space are identically zero since all vectors are parallel to each other.

Continuing with the Yang-Mills [5] –Shaw [6] extension of Maxwell electrodynamics, whose solution was obtained by C.R. Lee [7] and S.K.Wong [8], we use their notations in our subsequent discussion. According to the Feynman-Dyson schema we add Wong's equations for the particle carrying the isotopic spin $I^a$, $a = 1,2,3$.

$$m\ddot{x}_j = F_j(x, \dot{x}, t); \ j = 1,2,3$$

$$\dot{I}_a + g\varepsilon_{abc} A_j^b I^c \dot{x}_j = 0; \quad a = 1,2,3; \quad j = 1,2,3 \tag{53}$$

(in the time axial gauge $A_0^a = 0$). $A_i^b$ are the vector potentials with space components $j = 1,2,3$ and isotopic spin components $a = 1,2,3$.

Commutation relations are now given by

$$[x_j, x_k] = 0$$

$$m[x_j, \dot{x}_k] = i\hbar \delta_{jk}$$

$$[I_a, I_b] = i\hbar \varepsilon_{abc} I_c \ ; \tag{54}$$

$$[x_j, I^a] = 0 \tag{55}$$

Particle motion is affected by the generalized external Lorentz force

$$F_j(x, \dot{x}, t) = gE_j(x, t) + g\varepsilon_{jkl} \dot{x}_k B_l(x, t) \tag{56}$$

where $E_j(x,t) \equiv E_j^a(x,t) I^a$ and $B_j(x,t) \equiv B_j^a(x,t) I^a$ are three-dimensional vectors both in space and in isotopic internal space of the particle. They are the solutions of classical Yang-Mills equations

$$\partial_i B_i^a + g\varepsilon^{abc} A_i^b B_i^c = 0 \tag{57}$$

$$\frac{\partial B_i^a}{\partial t} + \varepsilon_{ijk}(\partial_j E_k^a + g\varepsilon^{abc} A_j^b E_k^c) = 0 \tag{58}$$

if the Weyl ordering prescription used.

The other two Yang-Mills equations define the charge and current densities:

$$\partial_i E_i^a + g\varepsilon^{abc} A_i^b E_i^c = \rho^a \tag{59}$$

$$-\frac{\partial E_i^a}{\partial t} + \varepsilon_{ijk}(\partial_j B_k^a + g\varepsilon^{abc} A_j^b B_k^c) = j_i^a \tag{60}$$

Thus again we have to deal with two sets of variables that describe the system dynamics.

Therefore the following system of equations suggests itself for further investigation:

$$[x_j, x_k] = 0$$

$$m[x_j, \dot{x}_k] = i\hbar \delta_{jk}$$

$$[I_a, I_b] = i\hbar f_{abc} I_c \ ; \quad a, b, c = 1, 2, ,, 7 \ ; \tag{61}$$

$$[x_j, I^a] = 0 \ ; \quad a = 1, 2, ,, 7; \quad j = 1, 2, 3 . \tag{62}$$

The generalized Lorentz force is expected to have the form

$$F_j(x,\dot{x},t) = gE_j(x,t) + g\varepsilon_{jkl}\dot{x}_k B_l(x,t) \tag{63}$$

where $E_j(x,t) \equiv E_j^a(x,t)I^a$ and $B_j(x,t) \equiv B_j^a(x,t)I^a$ are three-dimensional vectors in outer space and seven-dimensional vectors in the particle inner space. They are the expected solutions of the following classical Yang-Mills equations:

$$\partial_i B_i^a + gf^{abc} A_i^b B_i^c = 0 \tag{64}$$

$$\frac{\partial B_i^a}{\partial t} + \varepsilon_{ijk}(\partial_j E_k^a + gf^{abc} A_j^b E_k^c) = 0 \tag{65}$$

Together with the properly defined charge and current density:

$$\partial_i E_i^a + gf^{abc} A_i^b E_i^c = \rho^a \tag{66}$$

$$-\frac{\partial E_i^a}{\partial t} + \varepsilon_{ijk}(\partial_j B_k^a + gf^{abc} A_j^b B_k^c) = j_i^a \tag{67}$$

Although deserving of attention, this option has not yet been treated in the literature. However, a relativistic and quantum version of the proposed theory should be developed.

**Conclusion**

The central point of the present discussion is connected to the possible role of composition algebras in current and future applications in physics. Here we consider only the closest neighborhoods to the standard complex Hilbert space in detail. The common feature of the schemas herein presented is that they provide rich structures, potentially containing the required symmetries for including both strong and gravitation interactions into the overall unification picture while use of composition algebras leads to severe limitations upon the dimensions of the inner and outer spaces. They dictate the mathematical operations allowed and the form of the coupling of external forces within the given physical system. However, in this paper the interaction fields were only treated classically. Much more work needs to be done in order to clarify the physical content of the suggested constructs.